\documentclass[aps,amsmath,amssymb,twocolumn,floatfix]{revtex4-2}

\usepackage{graphicx}
\usepackage{dcolumn}
\usepackage{bm}

\DeclareUnicodeCharacter{2212}{-}

\begin{document}

\title{The influence of crystalline electric field on the magnetic properties of CeCd$_{3}X_{3}$~($X$~=~P~and~As) }
\author{Obinna P. Uzoh$^1$, Suyoung Kim$^1$, and Eundeok Mun$^{1,2}$}

\affiliation{$^1$ Department of Physics, Simon Fraser University, 8888 University Drive, Burnaby, B.C. Canada}
\affiliation{$^2$ Center for Quantum Materials and Superconductivity (CQMS), Sungkyunkwan University, Suwon 16419, South Korea}

\begin{abstract}
CeCd$_3$P$_3$ and CeCd$_3$As$_3$ compounds adopt the hexagonal ScAl$_3$C$_3$-type structure, where magnetic Ce ions on a triangular lattice order antiferromagnetically below $T_\text{N} \sim$0.42~K. Their crystalline electric field (CEF) level scheme has been determined by fitting magnetic susceptibility curves, magnetization isotherms, and Schottky anomalies in specific heat. The calculated results, incorporating the CEF excitation, Zeeman splitting, and molecular field, are in good agreement with the experimental data. The CEF model, with Ce$^{3+}$ ions in a trigonal symmetry, explains the strong easy-plane magnetic anisotropy that has been observed in this family of materials. A detailed examination of the CEF parameters suggests that the fourth order CEF parameter $B_{4}^{3}$ is responsible for the strong CEF induced magnetocrystalline anisotropy, with a large $ab$-plane moment and a small $c$-axis moment. The reliability of our CEF analysis is assessed by comparing the current study with earlier reports of CeCd$_{3}$As$_{3}$. For both CeCd$_{3}X_{3}$ ($X$ = P and As) compounds, less than 40 \% of $R\ln(2)$ magnetic entropy is recovered by $T_\text{N}$ and full $R\ln(2)$ entropy is achieved at the Weiss temperature $\theta_{p}$. Although the observed magnetic entropy is reminiscent of delocalized 4$f$-electron magnetism with significant Kondo screening, the electrical resistivity of these compounds follows a typical metallic behavior. Measurements of thermoelectric power further validate the absence of Kondo contribution in CeCd$_{3}X_{3}$.
\end{abstract}

\maketitle

\section{Introduction}

Several $f$-electron materials with triangular lattices (TL) have shown rich phenomena, where the spin-orbit coupling enhances quantum fluctuations due to the highly anisotropic interactions between 4$f$ moments~\cite{Hu2015, Li2016, Iqbal2016, Gong2017, Zhu2018, Rau2018}. For example, a spin liquid state has been proposed in insulating YbMgGaO$_{4}$~\cite{Li2015, Li2015PRL, Shen2016, Li2016YbMgGaO4, Paddison2017} and NaYbS$_{2}$\cite{Baenitz2018, Sichelschmidt2019}. The low carrier density system YbAl$_{3}$C$_{3}$ shows a gap in the magnetic excitation spectrum due to the dimerization of the $f$ electrons in Yb$^{3+}$ pairs~\cite{Ochiai2007, Kato2008, Ochiai2010, Hara2012}. Of interest is the easy-plane antiferromagnets CeCd$_3X_3$ ($X$ = P and As), a new class of TL system, with a low antiferromagnetic ordering temperature and extremely low carrier density~\cite{Liu2016, Dunsiger2020, Avers2021, Higuchi2016, Lee2019}.

\begin{figure}
\includegraphics[width=1\linewidth]{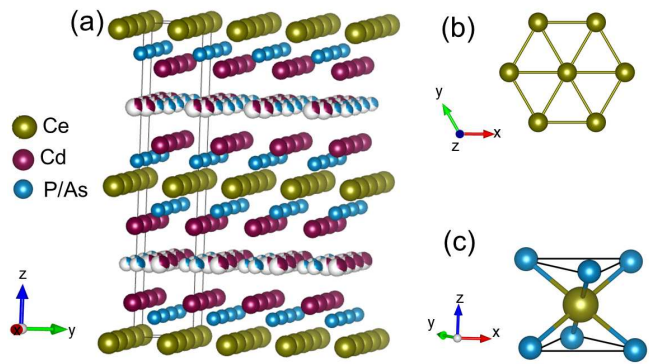}
\caption{(a) Crystal structure of the hexagonal CeCd$_3X_3$ ($X=$ P and As). (b) Ce atoms forming two-dimensional triangular lattices. (c) Ce atom in a trigonal ($D_{3d}$) point symmetry surrounded by P/As atoms.}
\label{FIG1}
\end{figure}

The family of compounds Ce$M_{3} X_{3}$ ($M$ = Al, Cd, and Zn, $X$ = C, P, and As) have been investigated for their robust ground state properties arising from the interplay between crystalline electric field (CEF) and magnetic exchange interaction on the triangular lattice~\cite{Ochiai2010, Liu2016, Dunsiger2020, Avers2021, Higuchi2016, Lee2019, Stoyko2011, Ochiai2021}. CeCd$_{3}$$X$$_{3}$ materials adopt the hexagonal ScAl$_{3}$C$_{3}$-type structure (space group $P6_{3}/mmc$) with the magnetic Ce$^{3+}$ ions having a  trigonal ($D_{3d}$) point symmetry as shown in Fig.~\ref{FIG1}(c)~\cite{Avers2021, Higuchi2016,  Lee2019, Stoyko2011}. In this crystal structure, the Ce layers are well separated by the Cd and $X$ atoms, forming a layered 2D TL in the $ab$-plane, as depicted in Figs.~\ref{FIG1}(a) and (b)~\cite{Avers2021, Higuchi2016,  Lee2019, Stoyko2011}. Thermodynamic and transport property measurements have characterized CeCd$_3$$X$$_3$ as an low carrier density system, with a strong easy plane magnetic anisotropy and antiferromagnetic ordering below $T_\text{N}$ $\sim$0.42~K~\cite{Dunsiger2020, Lee2019}. The low carrier density means there are insufficient charge carriers to screen all local moments and to mediate Ruderman-Kittel-Kasuya-Yosida (RKKY) exchange interaction between moments. In this regard, both RKKY and Kondo interactions are expected to be weakened in these compounds. We note that the electrical resistivity measurement on CeCd$_3$As$_3$ grown by chemical vapor transport shows a semiconductor-like enhancement as decreasing temperature ~\cite{Avers2021}, while the flux grown CeCd$_3$As$_3$ sample is metallic~\cite{Dunsiger2020}. At the magnetic ordering temperature, roughly 40\% of $R\ln(2)$ entropy is recovered. This implies a doublet ground state resulting from CEF splitting of localized Ce ion energy levels. The highly enhanced specific heat below 10~K and the reduced magnetic entropy at $T_\text{N}$ are reminiscent of Kondo lattice materials. However, the electrical resistivity of CeCd$_{3}X_{3}$ shows no maximum or logarithmic upturns resulting from the Kondo scattering of conduction electrons from magnetic Ce ions. The resistivity of CeCd$_{3}X_{3}$ is the same as that of LaCd$_{3}X_{3}$, implying an absence of Kondo screening in these materials~\cite{Dunsiger2020, Lee2019}.

Here, we use the CEF scheme to clarify the anisotropic magnetic properties of CeCd$_3X_{3}$ ($X$ = P and As), where the CEF analysis is performed using the $PyCrystalField$ package~\cite{Scheie2021}. For CeCd$_{3}$As$_{3}$, the CEF analyses of two previous independent studies~\cite{Avers2021, Banda2018}, carried out only considering magnetic susceptibility and magnetization, have shown discrepancies in their energy level splittings and first excited state wave functions. Thus, we extend the CEF analysis to specific heat data to resolve these discrepancies. In contrast to CeCd$_{3}$As$_{3}$, no CEF analysis has been carried out for CeCd$_{3}$P$_{3}$. We show that the easy plane magnetic anisotropy observed in these compounds can be explained by the strong CEF acting on Ce$^{3+}$ ions. In addition, we provide evidence of the lacking Kondo screening in these materials by way of thermoelectric power measurements.

\section{Experimental}

For this work, single crystals of  LaCd$_3X_3$ and CeCd$_3X_3$ ($X$ = P and As) were grown by high temperature ternary melt~\cite{Canfield1992, Lee2019, Dunsiger2020}. Magnetization measurements as a function of temperature and magnetic field were performed in a QD MPMS. Specific heat of these compounds was measured in QD PPMS. The obtained results are consistent with earlier reports~\cite{Dunsiger2020, Lee2019}. Thermoelectric power measurements were performed in a home made two thermometer and one heater setup. Detailed studies of thermodynamic and transport properties of these compounds are presented in Ref.~\cite{Dunsiger2020, Lee2019}.

\section{Results \& Discussion}

\begin{figure}
\includegraphics[width=1\linewidth]{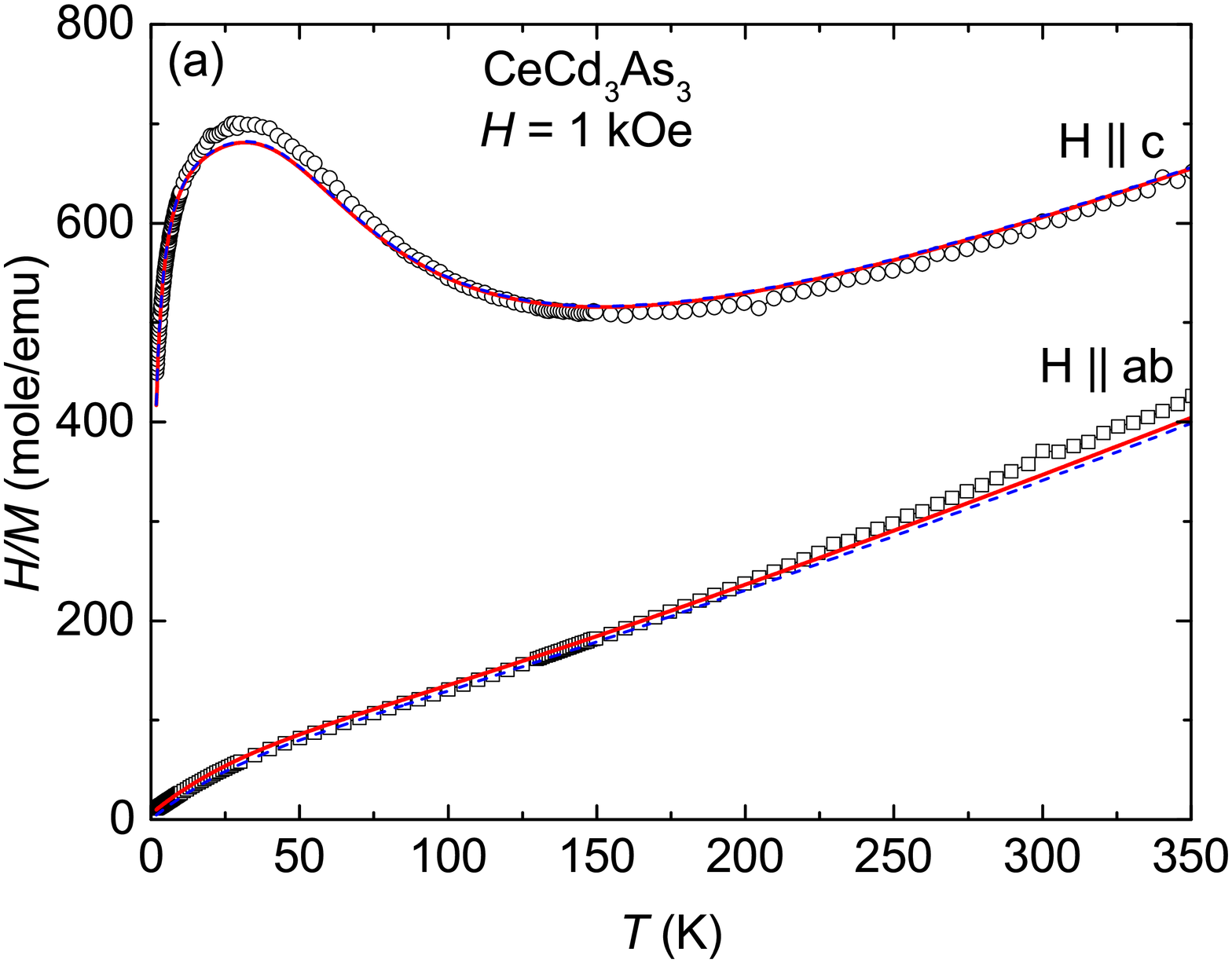}
\includegraphics[width=1\linewidth]{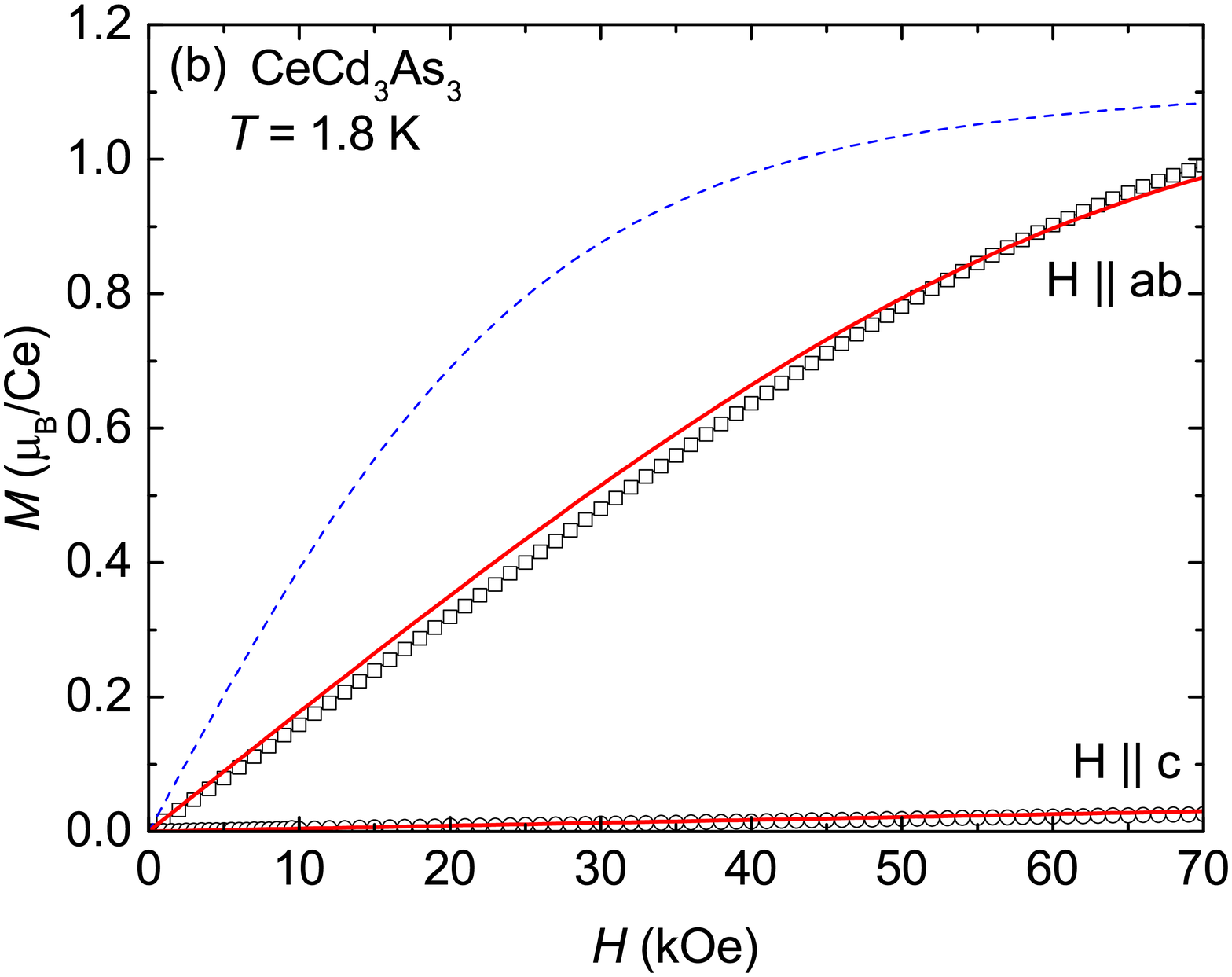}
\includegraphics[width=1\linewidth]{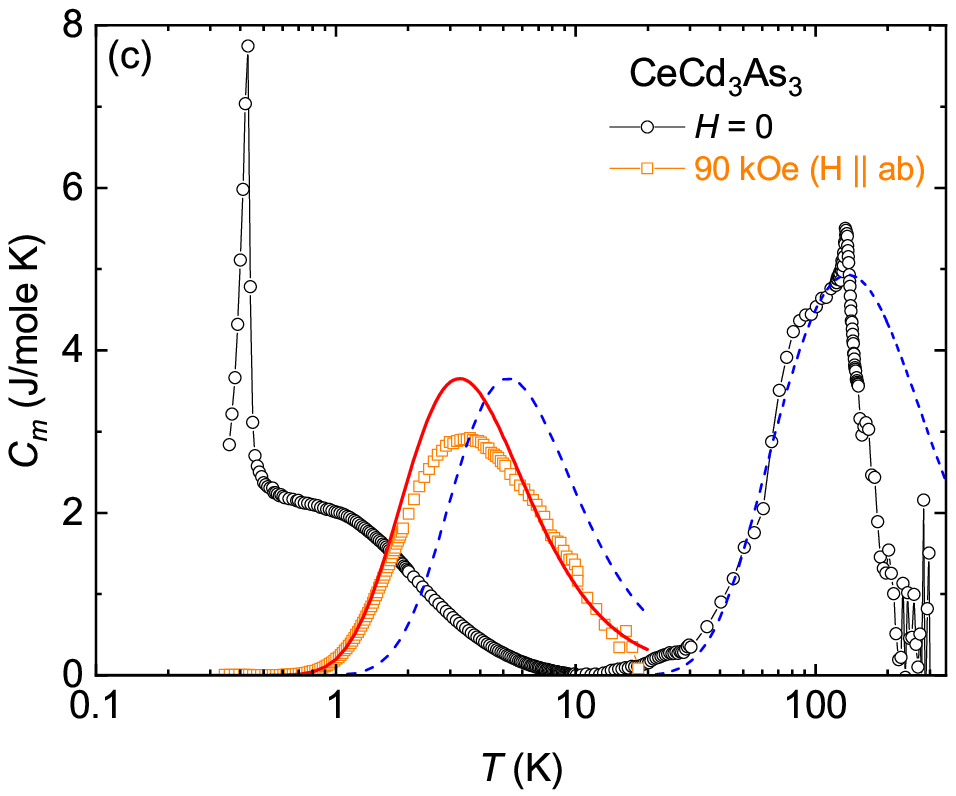}
\caption{(a) Inverse magnetic susceptibility curves of CeCd$_3$As$_3$, taken at $H$ = 1 kOe for both $H \parallel ab$ and $H \parallel c$. (b) Magnetization isotherms at 1.8~K for both $H \parallel ab$ and $H \parallel c$. (c) Magnetic part of the specific heat at $H$ = 0 and 90~kOe along $H \parallel ab$. Open symbols are experimental data. Dashed lines represent CEF fits with no molecular field contribution. Solid lines are CEF fits with molecular field terms $\lambda_{ab}~=~−5.7$~mole/emu and $\lambda_{c}~=~1.0$~mole/emu.}
\label{FIG2}
\end{figure}

\begin{table}
\caption{CeCd$_3$As$_3$: CEF parameters, energy eigenvalues, eigenstates, and molecular field parameters.}
\begin{tabular}{p{1.2cm}cccccc}
\hline
\hline
\multicolumn{7}{l}{CEF and molecular field parameters}\tabularnewline
\hline
&$B_2^0$ (K) & $B_4^0$ (K) & $B_4^3$ (K) & $\lambda_i$ (mole/emu)  \tabularnewline
&18.55 &  -0.08 & 23.02 & $\lambda_{ab} = -5.7$, \tabularnewline
&&&& $\lambda_c = 1.0$ \tabularnewline

\multicolumn{7}{l}{Energy eigenvalues and eigenstates}\tabularnewline
\hline
E (K) &$| -\frac{5}{2}\rangle$ & $| -\frac{3}{2}\rangle$ & $| -\frac{1}{2}\rangle$ & $| \frac{1}{2}\rangle$ & $| \frac{3}{2}\rangle$ & $| \frac{5}{2}\rangle$ \tabularnewline
\hline 
0 & 0.0 & 0.0 & $-$0.898 & 0.0 & 0.0 & 0.440 \tabularnewline
0 & 0.440 & 0.0 & 0.0 & 0.898 & 0.0 & 0.0 \tabularnewline
242 & 0.0 & -1.0 & 0.0 & 0.0 & 0.0 & 0.0 \tabularnewline
242 & 0.0 & 0.0 & 0.0 & 0.0 & -1.0 & 0.0 \tabularnewline
553 & -0.898 & 0.0 & 0.0 & 0.440 & 0.0 & 0.0 \tabularnewline
553 & 0.0 & 0.0 & 0.440 & 0.0 & 0.0 & 0.898 \tabularnewline
\hline
\hline
\end{tabular}
\label{Table1}
\end{table}

For CeCd$_3$$X$$_3$ systems, Ce atoms occupy a single site at the 2$a$ Wyckoff position that has a trigonal point symmetry, shown in Fig.~\ref{FIG1}(c). In this point symmetry, the CEF Hamiltonian requires only three parameters~\cite{Huang2015, Shin2020}: ${\cal H}_{CEF} = B^{0}_{2}O^{0}_{2} + B^{0}_{4}O^{0}_{4} + B^{3}_{4}O^{3}_{4}$, where $O^{m}_{n}$ are the Stevens operators  and $B^{m}_{n}$ are known as the CEF parameters~\cite{Stevens1952, Hutchings1964}. It has to be noted here that thermodynamic and transport property measurements of these compounds indicate a possible structural phase transition below 132~K~\cite{Ochiai2010, Dunsiger2020, Lee2019}. Additional CEF parameters below the transition temperature are probably required to account for the change in the local environment of the Ce atoms. However, the magnetic susceptibility curves show a smooth evolution through the transition temperature. Therefore, we assume in our CEF analysis that the Ce ions have the same trigonal symmetry below the transition temperature. To account for the Zeeman effect and the interaction between magnetic ions, the eigenvalues and eigenfunctions are determined by diagonalizing the total Hamiltonian: ${\cal H} = {\cal H}_{CEF} - g_{J} \mu_{B} J_{i} (H_{i} + \lambda_{i} M_{i})$ ($i$ = $x$, $y$, and $z$), where $g_J$ is the Land\'{e} $g$-factor, $\mu_B$ is the Bohr magneton, $H_i$ is the applied field, $J_i$ is the angular momentum operator, and $M_i$ is the magnetization. The second term is the Zeeman contribution and the third term represents the effective molecular interactions $\lambda_{i}$.

\begin{figure}
\includegraphics[width=1\linewidth]{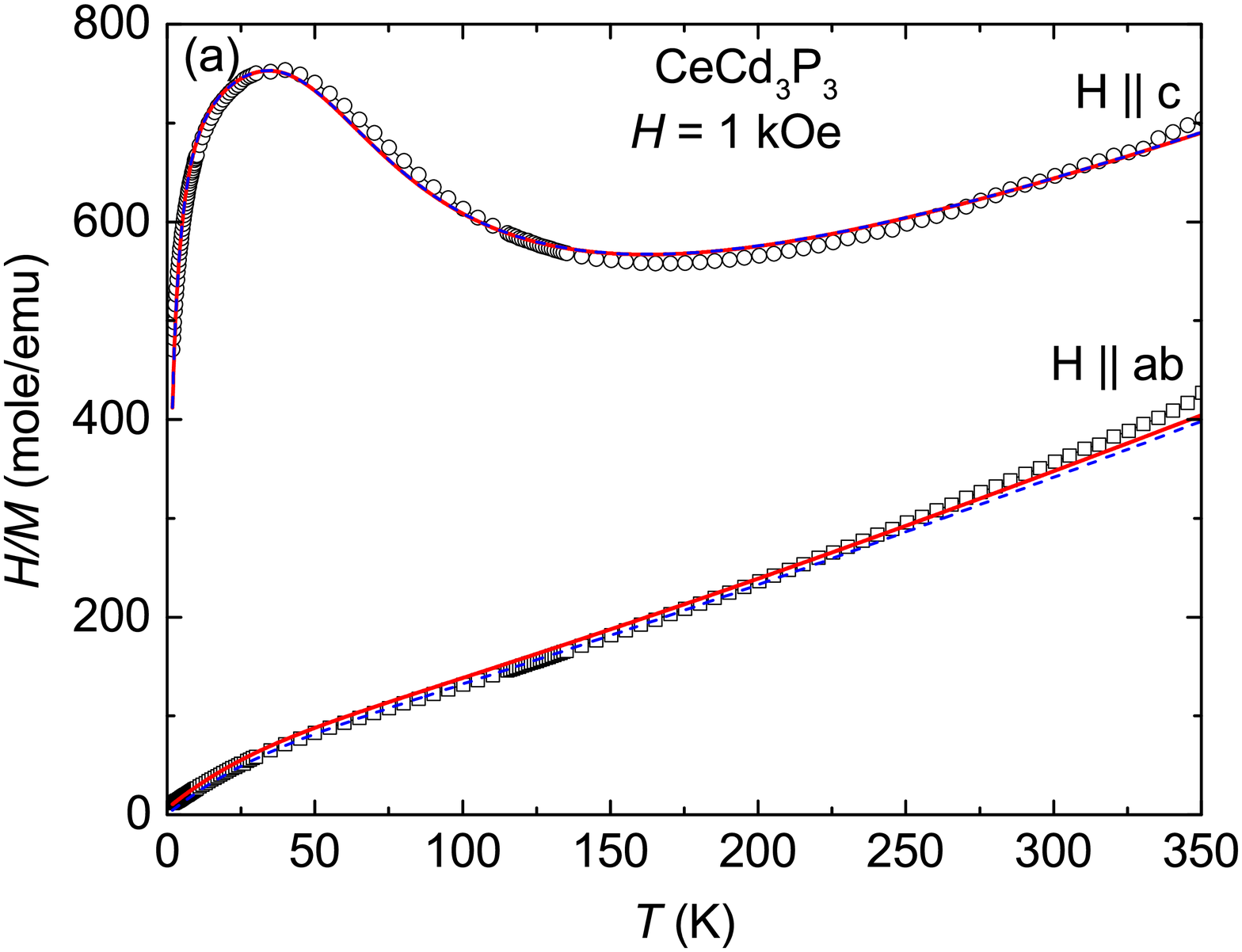}
\includegraphics[width=1\linewidth]{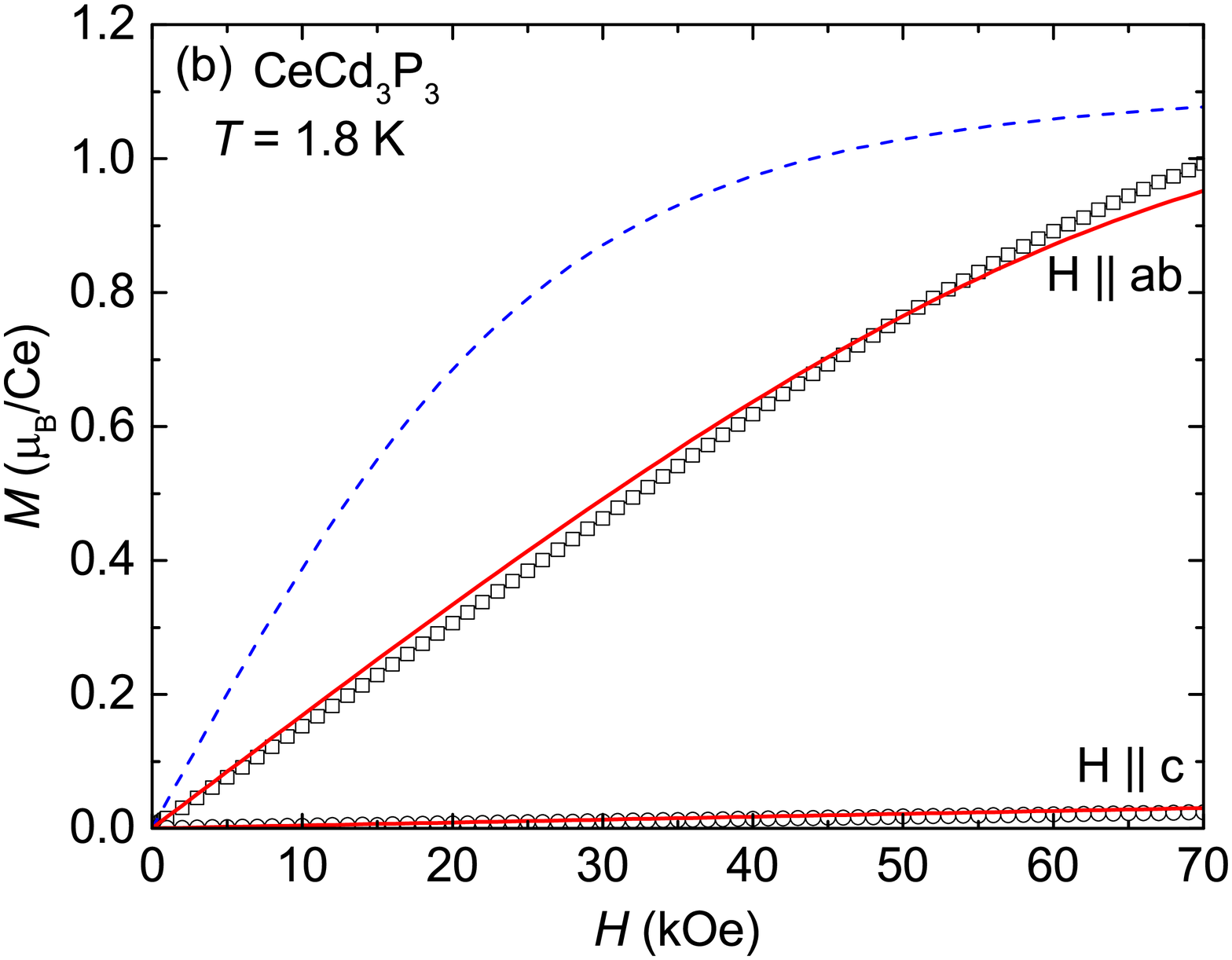}
\includegraphics[width=1\linewidth]{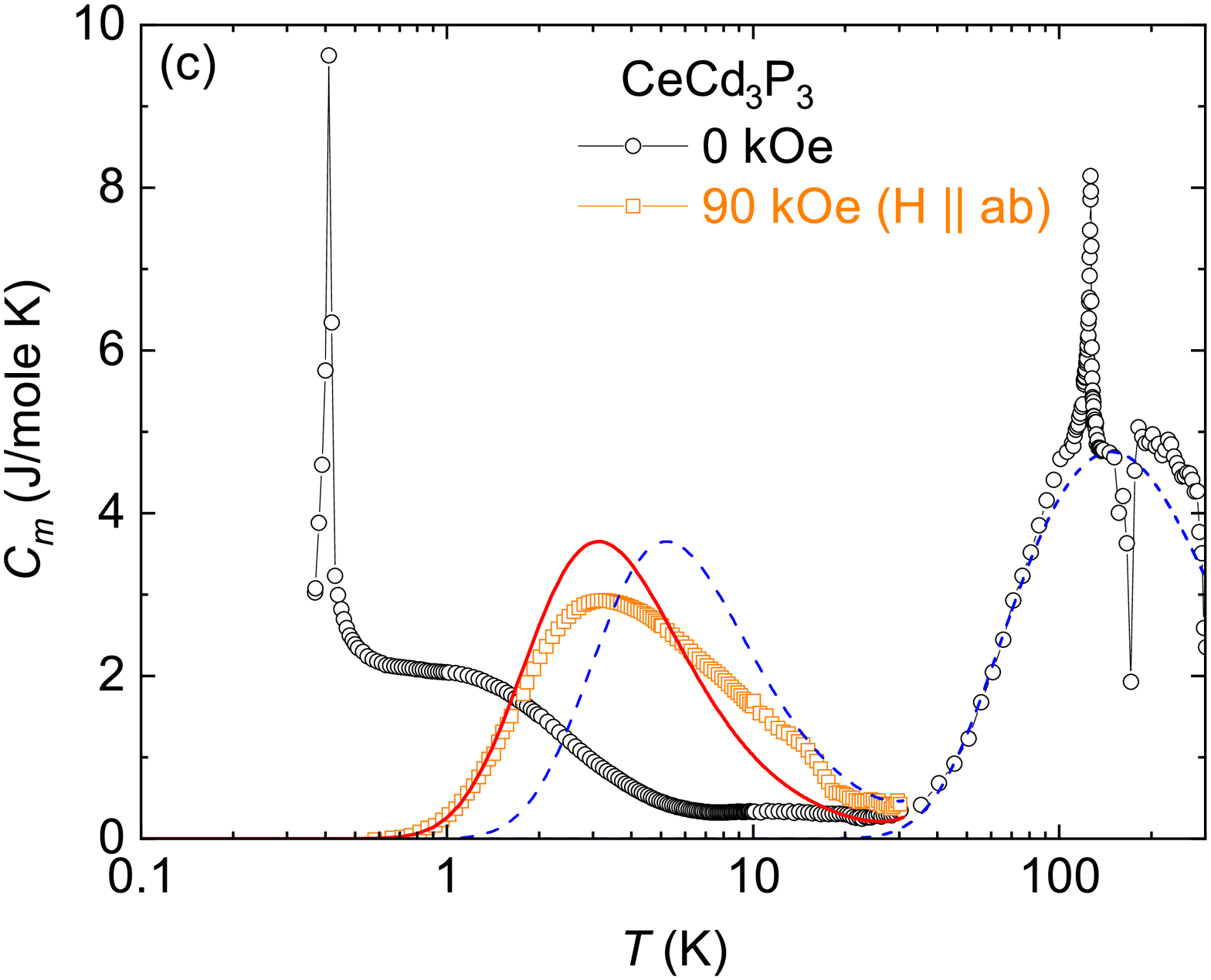}
\caption{(a) Inverse magnetic susceptibility curves of CeCd$_3$P$_3$, taken at $H$ = 1 kOe for both $H \parallel ab$ and $H \parallel c$. (b) Magnetization isotherms at 1.8~K for both $H \parallel ab$ and $H \parallel c$. (c) Magnetic part of the specific heat at $H$ = 0 and 90~kOe along $H \parallel ab$. Open symbols are experimental data. Dashed lines represent CEF fits with no molecular field contribution. Solid lines are CEF fits with molecular field terms $\lambda_{ab}~=~−6.2$~mole/emu and $\lambda_{c}~=~0.3$~mole/emu.}
\label{FIG3}
\end{figure}

The point charge CEF model assumes that the magnetic ions are well localized with a stable valence state. The valence state of CeCd$_3$$X$$_3$ can be deduced from the effective moment of Ce ions. Figures~\ref{FIG2}(a) and \ref{FIG3}(a) show inverse magnetic susceptibility, $\chi^{-1} = H/M$, curves of CeCd$_3$As$_3$ and CeCd$_3$P$_3$, respectively. At sufficiently high temperatures, magnetic susceptibility curves follow the Curie-Weiss (CW) behavior: $\chi(T) = C/(T - \theta_{p})$, where $C$ is the Curie constant and $\theta_{p}$ is the Weiss temperature. The effective moments of CeCd$_3$P$_3$ and CeCd$_3$As$_3$ are estimated by applying the CW law to the polycrystalline averaged magnetic susceptibility to be $\mu_{eff} = 2.51~\mu_B$ and $\mu_{eff} = 2.54~\mu_B$, respectively; which agree very well with the theoretical value $\mu_{eff} = 2.54~\mu_B$. From the effective moment values, it is reasonable to assume that Ce ions in these compounds are well localized with 3+ valence state. As shown in Ref.~\cite{Klingner2011, Banda2018} the $B^0_2$ parameter mainly gives a measure of the magnetocrystalline anisotropy and can be expressed in terms of the Weiss temperatures~\cite{Wang1971, Bowden1971, Shohata1977}. From the Curie-Weiss fit (not shown), anisotropic Weiss temperatures of CeCd$_3$As$_3$ are estimated to be $\theta_{ab} \sim 9.3$~K  and $\theta_{c} = -283$~K. For CeCd$_3$P$_3$, $\theta_{ab} = 9.3$~K and $\theta_{c} = -248$~K are obtained. The obtained $\theta_{p}$ values are consistent with earlier reports~\cite{Dunsiger2020, Lee2019}. These values are used to estimate the leading CEF parameters: $B^0_{2} \sim 30.5$~K for CeCd$_3$As$_3$ and $B^0_2 \sim 26.7$~K for CeCd$_3$P$_3$.

Figure~\ref{FIG2} displays anisotropic magnetic susceptibility curves, magnetization isotherms, and specific heat of CeCd$_3$As$_3$ with the results of CEF analysis. Table \ref{Table1} summarizes the obtained CEF parameters, including molecular field parameters, and the energy eigenstates and eigenvalues. The $2J + 1$ degenerate levels for $J = 5/2$ of Ce$^{3+}$ split into three Kramers doublets with energy level splittings $\Delta_1$ = 242~K and $\Delta_2$ = 553~K. The ground state and second excited state are in a mixture of $\vert \pm5/2 \rangle$ and $\vert \mp1/2 \rangle$ states. However, the first excited state is a pure $\vert \pm3/2 \rangle$ state. 

First, we discuss the CEF effects on thermodynamic properties of CeCd$_3$As$_3$ without the molecular field contribution. In Fig.~\ref{FIG2}, blue dashed-lines represent the CEF evaluations with $\lambda_{i}$ = 0. The CEF fit generally agrees with the inverse magnetic susceptibility as the large anisotropy between crystallographic directions is captured. For $H \parallel c$ the broad hump around 25 K is well reproduced. The CEF model aligns very well with the magnetization isotherm, $M(H)$, for $H \parallel c$ at 1.8 K, whereas there is an inconsistency between the CEF calculation and $M(H)$ for $H \parallel ab$, as shown in Fig.~\ref{FIG2}(b). The magnetic contribution to the specific heat, $C_m$, at $H$ = 0 and 90~kOe is presented in Fig.~\ref{FIG2}(c), where $C_m$ is estimated by subtracting the specific heat of non-magnetic analog LaCd$_3$As$_3$. In zero field, the broad maximum near 130~K in $C_m$ can be reproduced by the CEF model. Note that the CEF calculation above the maximum cannot capture the experimental data, which is caused by the very large subtraction error as explained in Ref.~\cite{Dunsiger2020}. Obviously, the sharp rise of $C_{m}$ below 10~K and the sharp peak at $T_\text{N}$ = 0.42~K~\cite{Dunsiger2020} are not captured by the CEF model. At $H$ = 90~kOe for $H \parallel ab$, the overall shape of $C_m$ is captured by the CEF model, but the maximum temperature is higher than that of experimental data.

The CEF model without $\lambda_{i}$ does not adequately reproduce $M(H)$ and $C_m$ data for $H \parallel ab$. In order to account for this mismatch, the molecular field interactions between Ce$^{3+}$ ions are incorporated. Red solid-lines in Fig.~\ref{FIG2} represent the CEF model in the presence of the molecular field terms: $\lambda_{ab} = -5.7$ mole/emu and $\lambda_{c} = 1$ mole/emu. The magnetic susceptibility curves with $\lambda_{i}$ show a slightly better agreement than that with $\lambda_{i} = 0$. However, as can be clearly seen in Fig.~\ref{FIG2}(b), the $M(H)$ curve for $H \parallel ab$ is well captured by the combination of the CEF and $\lambda_{i}$. Moreover, although the absolute value of the maximum in $C_m$ at 90 kOe is slightly higher, the position of the maximum temperature is well reproduced by introducing $\lambda_{i}$. These results point to the importance of including the exchange interactions between Ce$^{3+}$ magnetic ions.

\begin{table}
\caption{CeCd$_3$P$_3$: CEF parameters, energy eigenvalues, eigenstates, and molecular field parameters.}
\begin{tabular}{p{1.2cm}cccccc}
\hline
\hline
\multicolumn{7}{l}{CEF and molecular field parameters}\tabularnewline
\hline
& $B_2^0$ (K) & $B_4^0$ (K) & $B_4^3$ (K) & $\lambda_i$ (mole/emu)  \tabularnewline
& 20.90 &  -0.03 & 26.00 & $\lambda_{ab} = -6.2$, \tabularnewline
&&&& $\lambda_c = 0.3$ \tabularnewline
\multicolumn{7}{l}{Energy eigenvalues and eigenstates}\tabularnewline
\hline
E (K) &$| -\frac{5}{2}\rangle$ & $| -\frac{3}{2}\rangle$ & $| -\frac{1}{2}\rangle$ & $| \frac{1}{2}\rangle$ & $| \frac{3}{2}\rangle$ & $| \frac{5}{2}\rangle$ \tabularnewline
\hline 
0 & 0.0 & 0.0 & $-$0.897 & 0.0 & 0.0 & 0.442 \tabularnewline
0 & 0.442 & 0.0 & 0.0 & 0.897 & 0.0 & 0.0 \tabularnewline
257 & 0.0 & -1.0 & 0.0 & 0.0 & 0.0 & 0.0 \tabularnewline
257 & 0.0 & 0.0 & 0.0 & 0.0 & -1.0 & 0.0 \tabularnewline
621 & -0.897 & 0.0 & 0.0 & 0.442 & 0.0 & 0.0 \tabularnewline
621 & 0.0 & 0.0 & 0.442 & 0.0 & 0.0 & 0.897 \tabularnewline
\hline
\hline
\end{tabular}
\label{Table2}
\end{table}

Now turning to the CEF analysis for CeCd$_3$P$_3$, following the same procedure applied to CeCd$_{3}$As$_{3}$. The isostructural compounds CeCd$_3$P$_3$ and CeCd$_3$As$_3$ show remarkably similar magnetic properties, implying that their local CEF environments are in close resemblance. Hence, it is expected that the CEF parameters will be quite similar for both compounds, giving rise to very similar CEF energy level splittings and eigenstates. Table~\ref{Table2} shows a summary of CEF fit results of CeCd$_{3}$P$_{3}$. As expected, the obtained CEF parameters, eigenstates, and eigenvalues for CeCd$_{3}$P$_{3}$ are very similar to those of CeCd$_{3}$As$_{3}$ (Table \ref{Table1}). The positive $B_2^0$ term indicates that the magnetization lies in easy plane as seen in Fig. \ref{FIG3}(b). The large $B_4^3$ term implies a mixed ground state with $\vert \pm 5/2 \rangle$ and $\vert \mp 1/2 \rangle$, just like the case for CeCd$_{3}$As$_{3}$. The first excited state is in a pure state of $\vert \pm 3/2 \rangle$ and the second excited state is in an admixture of $\vert \pm 5/2 \rangle$ and $\vert \mp 1/2 \rangle$ states. The energy level splittings correspond to 257~K for the first excited state and 621~K for the second excited state.

Figure~\ref{FIG3} shows magnetic susceptibility, magnetization, and specific heat curves of CeCd$_{3}$P$_{3}$ together with the CEF calculations. In the absence of molecular field terms, the CEF calculations (dashed-lines) agree very well with the experimental $H/M$ curves, as shown in Fig.~\ref{FIG3}(a). Also, in zero field $C_m$ agrees with the CEF fit at high temperatures, as shown in Fig.~\ref{FIG3}(c), implying the high temperature broad maximum (Schottky anomaly) is due to the CEF effects. The inclusion of the molecular field terms greatly alters the $M(H)$ curve calculation for $H \parallel ab$ and adequately captures the experimental $M(H)$ data. In addition, the CEF calculation with molecular field terms properly captures the low temperature maximum in $C_m$ at $H$ = 90~kOe. Therefore, the CEF model in the present of the molecular field interaction should be used to describe the experimental data for CeCd$_3$P$_3$. As expected from the layered crystal structure, the absolute value of $\lambda_{ab}$ is greater than $\lambda_c$, implying a strong exchange interaction in the $ab$-plane.

The obtained CEF parameters of CeCd$_3$P$_3$ are slightly larger than those of CeCd$_3$As$_3$. The slight difference in the values of their CEF parameters is quite surprising. When the distance between Ce and nearest As/P atoms is considered, the difference between Ce-As distance (3.05~\text{\AA}) and Ce-P distance (2.96~\text{\AA}) is about $\sim$9~pm \cite{Dunsiger2020, Avers2021, Lee2019}, which might not be large enough to cause different CEF energy level splittings. Note that the validity of the CEF Hamiltonian must be verified below the (structural) phase transition temperature $T_{s}$ = 127~K for CeCd$_{3}$P$_{3}$~\cite{Lee2019} and $T_{s}$ = 136~K for CeCd$_{3}$As$_{3}$~\cite{Dunsiger2020}.

\begin{figure}
\includegraphics[width=1\linewidth]{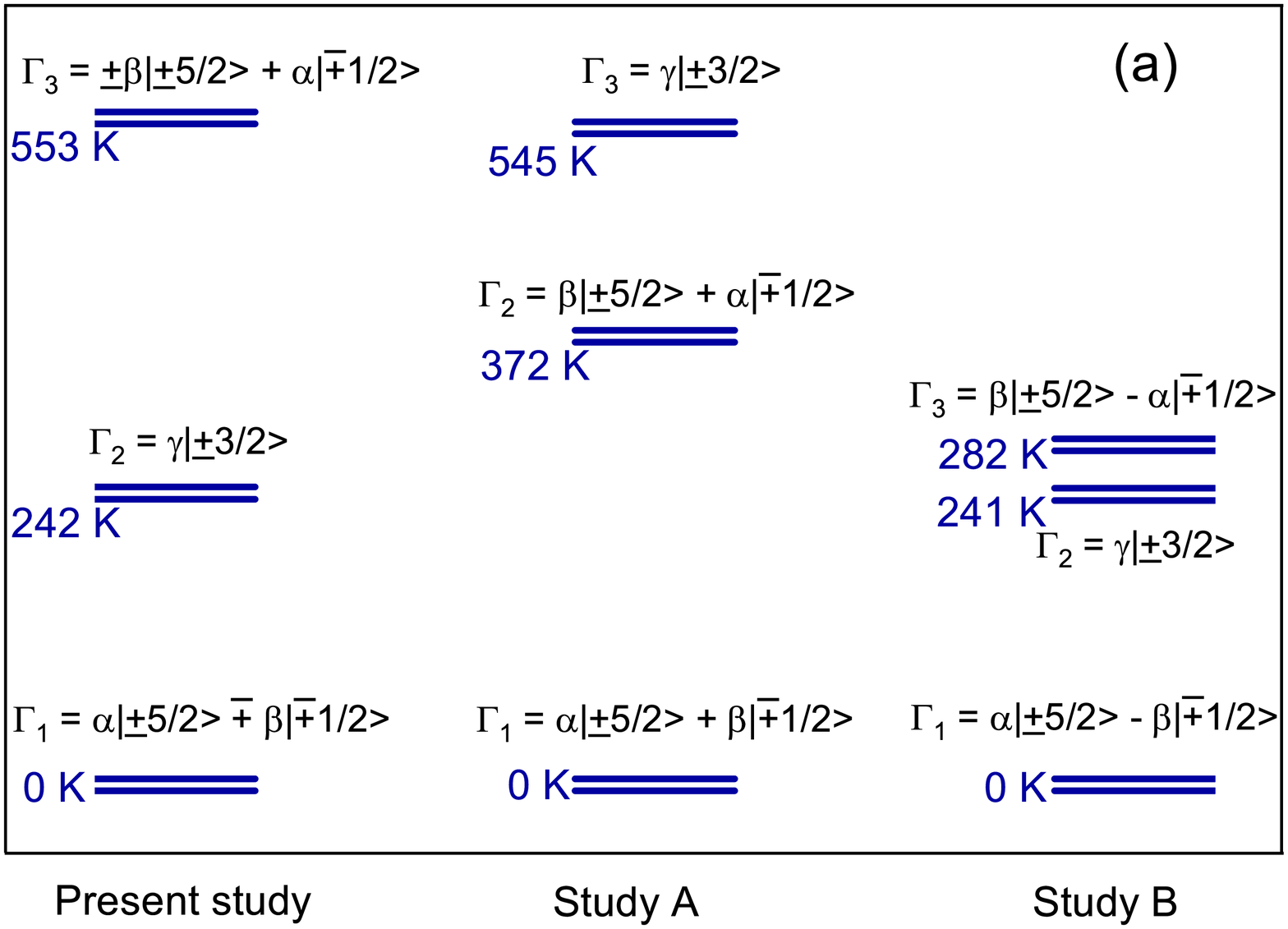}
\includegraphics[width=1\linewidth]{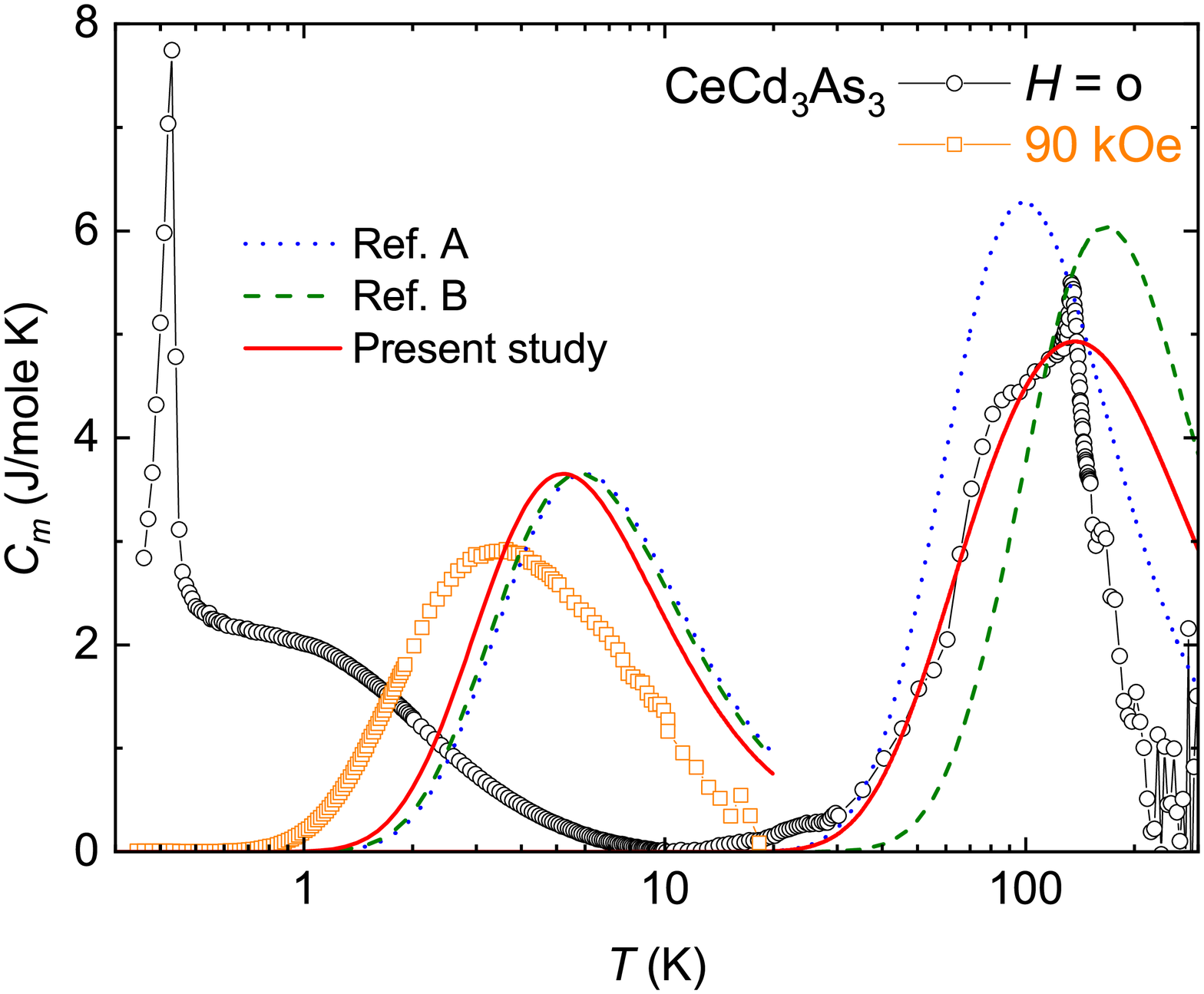}
\caption{CEF profiles of CeCd$_3$As$_3$ for three independent studies. Study A is from Ref.~\cite{Avers2021} and Study B is from Ref.~\cite{Banda2018}. (a) CEF energy level splittings and eigenstates. Present study: $\alpha~=~0.44$, $\beta~=~0.90$, and $\gamma~=~-1$. Study A: $\alpha~=~0.32$, $\beta~=~0.95$, and $\gamma~=~1$. Study B: $\alpha~=~0.28$, $\beta~=~0.96$, and $\gamma~=~1$. (b) Comparison between $C_m$ of present study and CEF calculations based on the three independent studies, where $\lambda_{i} = 0$.}
\label{FIG4}
\end{figure}

In Fig.~\ref{FIG4}(a) we compare the energy level splittings and eigenfunctions for three independent CEF analyses of CeCd$_{3}$As$_{3}$. For all three studies the CEF parameters are obtained from fits to the magnetic susceptibility curves. In this study, the Hamiltonian includes the molecular field term. In Ref.~\cite{Avers2021} (Study A), the CEF parameters are obtained in the presence of antiferromagnetic exchange interaction based on a mean-field approach. In Ref.~\cite{Banda2018} (Study B), the CEF parameters are evaluated by fitting the magnetic susceptibility curves to only ${\cal H}_{CEF}$ (without any interaction term).

For all three studies, the ground state ($\Gamma_{1}$) is in a mixed state of $\vert \pm 5/2 \rangle$ and $ \vert \mp 1/2 \rangle$ with a higher probability in the $\vert \mp 1/2 \rangle$ state. This requires a mixing angle $\theta$ in the wave function: $\cos(\theta) \vert \pm 5/2 \rangle + \sin(\theta)\vert \mp 1/2 \rangle$. The obtained mixing angle is roughly similar for all three studies: 64$^\circ$ in this study, 72$^\circ$ in Ref.~\cite{Avers2021}, and 74$^\circ$ in Ref.~\cite{Banda2018}. Unlike the ground state, the excited states $\Gamma_{2}$ and $\Gamma_{3}$ do not entirely agree for all three studies. In Ref.~\cite{Avers2021} there is a clear swap between $\Gamma_{2}$ and $\Gamma_{3}$ wave functions (Study A). This is caused by the relatively high $\vert B_{4}^{0} \vert$ value (= 1.4~K). In fact, we confirmed that any $\vert B_{4}^{0} \vert > 0.7$~K would result in the $\Gamma_{2}$ being in a mixed state and the $\Gamma_{3}$ being in a pure state, just as the case in Ref.~\cite{Avers2021}. Another important distinction among these three studies is in the energy level splittings. The energy eigenvalue for the second excited state in Ref.~\cite{Banda2018} (Study B) is roughly two times smaller than that of the other two studies. However, the energy level splitting from the ground to the first excited state ($\sim$240~K) is comparable for all studies, clearly indicating that the ground state is well isolated from the excited states.

Figure~\ref{FIG4} (b) shows the magnetic part of the specific heat, together with the calculated specific heat curves by using CEF parameters. Solid lines, dashed lines, and dotted lines are CEF calculations from the present study, Ref.~\cite{Avers2021}, and Ref.~\cite{Banda2018}, respectively. As shown in the figure, when subtle differences are ignored, the high temperature maximum is captured by all three calculated curves. This implies that the high temperature maximum in $C_{m}$ is due to the CEF effect and the ground state doublet is well isolated from the excited states. When measurement uncertainty and different sample quality are considered, the best CEF parameters among three parameter sets cannot be selected from the comparison with $C_{m}$. Therefore, further measurements such as inelastic neutron scattering with Cd isotopes are necessary to distinctly specify the best CEF parameters in this system. In addition, the CEF scheme can be determined by optical spectroscopy techniques such as Raman scattering. Note that CEF parameters evaluated by the three independent studies provide a qualitatively good description of the experimental $M/H$ curves.

The significance of $B_{4}^{3}$ CEF parameter has been observed in Ce-based antiferromagnets such as CeAuSn, CeIr$_3$Ge$_7$, and CeCd$_{3}X_{3}$, where Ce ions are in a trigonal environment~\cite{Adroja1997, Huang2015, Banda2018}. These compounds indicate a large magnetic anisotropy with the $ab$-plane being the magnetic easy plane, which can be qualitatively explained by the CEF effect. A detailed CEF analysis based on both the magnetic susceptibility and inelastic neutron scattering data of hexagonal CeAuSn indicates a mixture of $\vert \pm 5/2\rangle$ and $\vert \mp 1/2 \rangle$ CEF ground state doublet, a pure $\vert \pm 3/2 \rangle$ doublet as the first excited state at $\sim$345~K, and a mixture of $\vert \pm 5/2\rangle$ and $\vert \mp 1/2 \rangle$ doublet as the second excited state at $\sim$440~K~\cite{Adroja1997, Huang2015}. The anisotropy ratio of magnetic susceptibility between $H \parallel ab$ and $H \parallel c$ is found to be $\chi_{ab}/\chi_{c} \sim 15$ near $T_{N}$. Both magnetic susceptibility~\cite{Huang2015} and neutron scattering ~\cite{Adroja1997} CEF evaluations clearly show a significant $B_4^3$ contribution ($\sim$19~K) with a large mixing angle, which is consistent with the CEF analysis of CeCd$_{3}X_{3}$. The anisotropy ratio of CeCd$_{3}X_{3}$ ($\chi_{ab}/\chi_{c} \sim 36$ at 1.8~K) is larger than that of CeAuSn because both $B_{2}^{0}$ and $B_{4}^{3}$ CEF parameters of CeCd$_{3}X_{3}$ are larger than those of CeAuSn. The CEF investigation of the rhombohedral CeIr$_3$Ge$_7$ compound shows a very similar CEF eigenstates and eigenvalues with those of other compounds. However, in CeIr$_3$Ge$_7$, the reported CEF parameters (especially the term $B_4^3 = 67.3$~K) are larger than that of other compounds, thus inducing a huge energy level splittings of 374~K and 1398~K.~\cite{Banda2018}. It has been suggested that the exceptionally large CEF splitting can be related to the contribution of the $5d$ ligands of Ir atoms.~\cite{Banda2018}. In addition, the CEF analysis on rhombohedral CePtAl$_4$Ge$_2$ antiferromagnet has also been conducted~\cite{Shin2020}. Unlike the above mentioned compounds, the ground state and the second excited state of CePtAl$_4$Ge$_2$ are not in a mixed configuration of $\vert 5/2\rangle$ and $\vert 1/2 \rangle$ states. The sign and magnitude of $B_2^0$ (= 13.26~K) and $B_4^0$ (= $-0.3$~K) CEF parameters in CePtAl$_4$Ge$_2$ are comparable to that of CeAuSn, CeIr$_3$Ge$_7$, and CeCd$_{3}$X$_{3}$ ($X$ = P and As). However, because the $B_4^3$ term responsible for mixing is exceptionally small in CePtAl$_4$Ge$_2$ system ($B_4^3 = 0 \pm 0.02$ K), the ground state and second excited state is in a pure  $\vert \pm 1/2 \rangle$ state and a  pure $\vert \mp 5/2 \rangle$ state, respectively. We found that the ground and excited states are without mixing for any $\vert B_4^3 \vert < 1.1$~K. The small value of $B_{2}^{0}$ and $B_{4}^{3}$ implies a relatively smaller magnetic anisotropy in CePtAl$_4$Ge$_2$, which is clearly reflected on its magnetic susceptibility data ($\chi_{ab}/\chi_{c} \sim 4$ near $T_{N}$)~\cite{Shin2020}.

\begin{figure}
\includegraphics[width=1\linewidth]{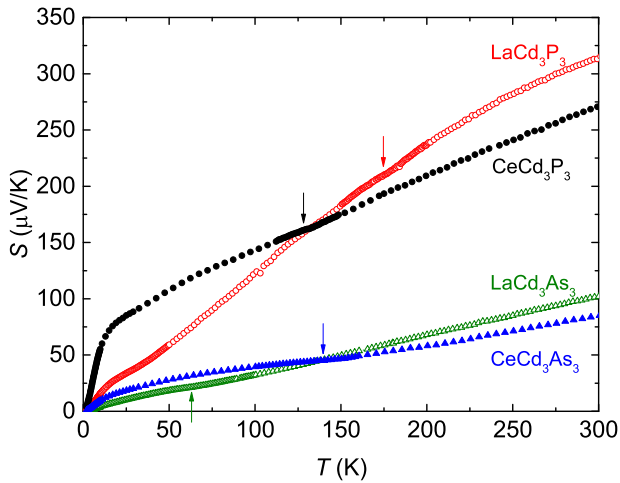}
\caption{Temperature dependence of the thermoelectric power, $S(T )$, of $R$Cd$_{3}$$X$$_{3}$ ($R$ = La and Ce, $X$ = P and As). Vertical arrows indicate the phase transition temperatures observed in electrical resistivity measurements \cite{Lee2019, Dunsiger2020}. }
\label{FIG5}
\end{figure}

Although our CEF analysis on CeCd$_{3}$$X$$_{3}$ ($X$ = P and As) provides a comprehensive picture at high temperatures, a number of unanswered questions remains at low temperatures. When the temperature is much lower than the CEF splitting, the lowest Kramers doublet is only relevant to explain the observed magnetic ordering at $T_\text{N} = 0.42$~K and upturn in $C_{m}$ below 10~K. It is obvious that the temperature dependence and absolute value of $C_{m}$ below 10~K cannot be explained by the CEF effects as shown in Figs.~\ref{FIG3}(c) and \ref{FIG4}(c). 

Since magnetization isotherms at $T$ = 1.8~K for both compounds are reproduced well by CEF calculation with the ground state wave functions, the reduced magnetization value cannot be associated with the Kondo screening. This is consistent with the electrical resistivity results of CeCd$_{3}$$X$$_{3}$~\cite{Dunsiger2020, Lee2019}. The absence of Kondo effect in CeCd$_{3}$$X$$_{3}$ is also confirmed from thermoelectric power (TEP) measurements, as shown in Fig.~\ref{FIG5}. The observed TEP value of LaCd$_{3}$$X$$_{3}$ is an order of magnitude higher than that of typical metals, implying low carrier density system. The temperature dependence of the TEP, $S(T)$, of both La- and Ce-based compounds shows a hump at low temperatures, which can be related to the phonon drag. In general, $S(T)$ of typical metals shows a maximum, corresponding to the phonon-drag effect, where the maximum is expected to be located between $\Theta_{D}/5$ and $\Theta_{D}/12$ \cite{Blatt1976}. Many Ce- and Yb-based Kondo lattice systems have shown that $S(T)$ indicates an extrema with enhanced value, corresponding to the Kondo effect in conjunction with CEF effect. The TEP of CeCd$_{3}$$X$$_{3}$ exhibits behavior similar to that of LaCd$_{3}$$X$$_{3}$, implying negligible Kondo contributions. Hence, as suggested in Refs.~\cite{Dunsiger2020, Lee2019}, the enhancement of the specific heat below 10~K is likely related to either the magnetic frustration in triangular lattices \cite{Dunsiger2020, Lee2019, Liu2016, Avers2021, Higuchi2016} or simply the magnetic fluctuations observed in insulating antiferromagnets \cite{Robinson1960}.

The electrical resistivity of CeCd$_{3}$$X$$_{3}$ compounds exhibits a metallic behavior, suggesting that Ruderman-Kittel-Kasuya-Yosida (RKKY) interaction may be responsible for the magnetic ordering. However, it would have to be mediated by an extremely small number of charge carriers \cite{Dunsiger2020, Lee2019}. It has been qualitatively shown that the magnetic ordering temperature of non-Kondo materials scale with the distance between Ce ions, where the larger the Ce-Ce distance results in a smaller ordering temperature~\cite{Rai2018}. For example, a non-Kondo metal CeIr$_{3}$Ge$_{7}$ orders at an extremely low temperature $T_\text{N} = 0.63$~K due to the large Ce-Ce distance ($\sim$6~\AA). When the Ce-Ce distance ($\sim$4~\AA) for CeCd$_{3}$$X$$_{3}$ is considered, the magnetic ordering temperature is significantly suppressed compared to that of other Ce-based non-Kondo systems. On the contrary, it has been suggested that the superexchange interaction in low carrier density YbAl$_{3}$C$_{3}$ compound becomes dominant instead of the RKKY interaction, where the carrier concentration is estimated to be $n \sim$0.01 per formula unit (f.u.)~\cite{Ochiai2007}. When the carrier concentrations of CeCd$_{3}$As$_{3}$ ($n \sim$ 0.003/f.u.)~\cite{Dunsiger2020} and CeCd$_{3}$P$_{3}$ ($n \sim$ 0.002/f.u.)~\cite{Lee2019} compounds are considered, it is reasonable to assume that the superexchange interaction may be responsible for the antiferromagnetic ordering below 0.42~K. In addition, the partial $H-T$ phase diagram of these compounds, especially the field-induced increase of $T_\text{N}$, is similar to that of 2D insulating triangular lattice systems with easy-plane anisotropy~\cite{Lee1986, Seabra2011}. Furthermore, the low temperature specific heat of CeCd$_{3}$As$_{3}$ has been explained by anisotropic exchange Hamiltonian for an insulating, layered triangular lattice~\cite{Avers2021}.

\section{Conclusion}

At high temperatures, the observed magnetic properties of CeCd$_{3}$$X$$_{3}$ triangular lattice compounds can be well understood by considering the CEF effects with molecular filed contributions. The large anisotropy in the magnetic susceptibility and magnetization and the electronic Schottky anomaly in the specific heat are explained by energy level splittings of the $J$ = 5/2 Hund's rule ground state of Ce$^{3+}$ ions into three doublets. The striking similarity of the CEF profile of both compounds implies a very close resemblance of their crystal field environment. Three independent CEF analyses on CeCd$_3$As$_3$ indicate inconsistent CEF profiles, requiring further studies such as inelastic neutron scattering. When the temperature is well below the CEF splitting, the well-isolated Kramers' doublet ground state is responsible for the antiferromagnetic ordering and the large enhancement of specific heat below 10~K. Further measurements such as magnetization, neutron scattering, and nuclear magnetic resonance are necessary to provide additional insight into the nature of magnetism below $T_\text{N}$ and the role of anisotropic exchange interactions in the triangular motif.

\begin{acknowledgments}
This work was supported by the Canada Research Chairs, Natural Sciences and Engineering Research Council of Canada, and Canada Foundation for Innovation program. EM was supported by the Korean Ministry of Science and ICT (No. 2021R1A2C2010925) and by BrainLink program funded by the Ministry of Science and ICT (2022H1D3A3A01077468) through the National Research Foundation of Korea.
\end{acknowledgments}

\end{document}